\documentclass[aps,prx,twocolumn,superscriptaddress,10pt]{revtex4-2}

\usepackage{amsmath}
\usepackage{amssymb}
\usepackage{amsfonts}
\usepackage{bm}
\usepackage{graphicx}
\usepackage{braket}
\usepackage{microtype}
\usepackage{dsfont}
\usepackage{booktabs}
\usepackage{multirow}

\makeatletter

\DeclareMathOperator{\tr}{\rm{tr}}

\newcommand{\jila}{\affiliation{JILA, National Institute of Standards and Technology, University of Colorado, Boulder, Colorado, USA}}
\newcommand{\ctqm}{\affiliation{Center for Theory of Quantum Matter, University of Colorado, Boulder, Colorado, USA}}
\newcommand{\nist}{\affiliation{Time and Frequency Division, National Institute of Standards and Technology, Boulder, Colorado, USA}}
\newcommand{\phys}{\affiliation{Department of Physics, University of Colorado, Boulder, Colorado, USA}}
\newcommand{\homer}{\affiliation{Homer L. Dodge Department of Physics and Astronomy, The University of Oklahoma, Norman, Oklahoma 73019, USA}}
\newcommand{\cqr}{\affiliation{Center for Quantum Research and Technology, The University of Oklahoma, Norman, Oklahoma 73019, USA}}

\usepackage{xcolor}
\definecolor{apsblue}{HTML}{006298}

\usepackage[unicode=true,
 bookmarks=false,
 breaklinks=false, colorlinks=true]
 {hyperref}
\hypersetup{
  linkcolor = apsblue,
  citecolor = apsblue,
  urlcolor = apsblue,
}

\begin{abstract}
Scaling trapped-ion quantum sensors from single ions to large ensembles is a key challenge for next-generation precision measurements. At the same time, many ion species of interest for optical clocks and tests of fundamental physics lack closed cycling transitions required for direct laser cooling and state detection. Collective quantum logic spectroscopy addresses both limitations by coupling an ensemble of sensor, or spectroscopy, ions to one or more logic ions that provide sympathetic cooling and state readout. Here, we establish the fundamental performance limits and operating regimes of this protocol, identifying how the interaction strength, interrogation time, and logic-ensemble size govern sensitivity, dynamic range, and robustness to experimental imperfections. We show that quantum-limited sensitivity can be retained even with a single logic ion, while increasing the number of logic ions substantially improves readout efficiency and robustness. Beyond precision metrology, the same collective interface enables many-body measurements relevant to quantum information processing, including parity measurements and stabilizer-like syndrome extraction. Our results establish collective quantum logic spectroscopy as a scalable framework for optical clocks, quantum-enhanced sensing, and trapped-ion quantum information processing.
\end{abstract}

\begin{document}

\title{Collective Quantum Logic Spectroscopy}

\author{Raphael Kaubruegger}\thanks{These authors contributed equally to this work.}\jila\phys
\author{Matthew Patkowski}\thanks{These authors contributed equally to this work.}\jila\phys
\author{Yicheng Zhang}\homer\cqr
\author{Robert J. Lewis-Swan}\homer\cqr
\author{David B. Hume}\nist\phys
\author{Ana Maria Rey}\jila\phys\ctqm

\maketitle

\begin{figure*}[t]
    \centering
    \includegraphics[]{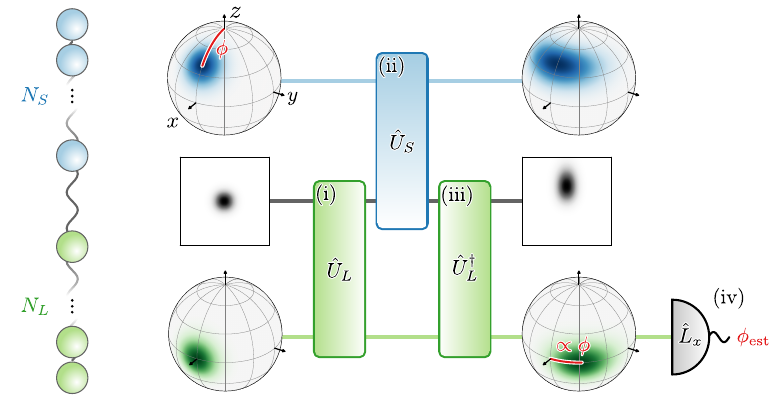}
    \caption{
    \textbf{Collective QLS protocol} 
$N_L$ logic ions  are initialized in a coherent spin state pointing along the $x$ axis, while $N_S$ spectroscopy ions  are initialized in a coherent spin state in which $\phi$ parametrizes the imbalance between atoms in the ground and excited states, as illustrated by their Wigner distributions on a generalized Bloch sphere. Both ensembles couple to a shared motional mode of the ion crystal, with creation operator $\hat a$, initially prepared in its ground state. The protocol proceeds by (i) entangling the logic ions with the mode through a spin-dependent displacement $\hat U_L$, (ii) imprinting the spectroscopy population onto the mode through $\hat U_S$, and (iii) reversing the logic--motion entangling operation with $\hat U_L^\dagger$. The Wigner distributions shown at the end of the sequence are obtained by tracing out the other two subsystems. The net effect is to map $\phi$ onto a collective rotation of the logic ensemble, which is then measured to estimate the parameter initially encoded in the spectroscopy ions.}
    \label{fig:qls_sequence}
\end{figure*}

Optical atomic clocks \cite{ludlow2015optical} based on single trapped ions have achieved extraordinary accuracy and stability \cite{brewer2019al+,marshall2025high}, enabling stringent tests of fundamental physics and powerful applications in timekeeping and precision measurement \cite{safronova2018search,kozlov2018highly}. Many species relevant for precision spectroscopy, including most atomic ions, molecular ions and highly charged ions, lack a convenient cycling transitions suitable for direct fluorescence detection \cite{wolf2016non,chou2017preparation,micke2020coherent,cornejo2021quantum}. Quantum logic spectroscopy (QLS) offers a powerful route around this limitation by coherently mapping otherwise inaccessible internal states onto logic ions that can be measured with high fidelity \cite{schmidt2005spectroscopy,hume2007high,sinhal2020quantum,chou2020frequency}.

A key next step is to extend these capabilities to multi-ion spectroscopy, where larger ensembles promise improved short-term stability \cite{herschbach2012linear,hausser2025in+} and, when combined with entanglement, operation beyond the standard quantum limit \cite{pezze2018quantum,schulte2020prospects,colombo2022entanglement,kaubruegger2025progress}. This extension is experimentally challenging because ground-state cooling, mode control, and individual addressing become increasingly demanding in larger ion crystals \cite{chen2020efficient,feng2020efficient}.

Several scalable readout paradigms have emerged to address this bottleneck. Algorithmic QLS combining multi-ion control with quantum-circuit concepts has been proposed to compress information about a large spectroscopy ensemble into a smaller logic ensemble through structured gate sequences \cite{Schulte2016quantum}. Separately, multi-ion swap gates have been demonstrated that transfer qubit excitations between species achieving high fidelity at temperatures above the ground state of ion motion~\cite{kienzler2020quantum}. Collective QLS instead uses collective spin--motion couplings to transfer global spectroscopy observables, most notably the ground-state population, onto a logic ensemble in parallel. In Ref.~\cite{cui2022scalable}, this approach was used to map the number of spectroscopy ions in the ground state onto a collective rotation of the logic ions, mediated by a shared bosonic mode.

While these developments establish the feasibility of scalable QLS, their implications for precision metrology and clock operation have remained largely unexplored. Here we show that collective QLS is not merely a scalable readout technique, but a tunable interface for extracting metrologically relevant information from many-body spectroscopy states. This capability applies broadly, from uncorrelated spin-coherent states to highly entangled states, and makes collective QLS a versatile readout tool rather than a protocol tied to a single sensing regime.

We analyze collective QLS beyond previously explored regimes, focusing on the estimation of a parameter encoded in the spectroscopy population imbalance after Ramsey or Rabi interrogation. This directly connects the protocol to the central metrological task of optical clocks. We determine how the interaction time with the shared mode controls information transfer, estimator variance, and dynamic range; how performance scales with the numbers of spectroscopy and logic ions, $N_S$ and $N_L$; and how realistic imperfections, including thermal phonon occupation, beyond-Lamb-Dicke corrections, and mode-frequency fluctuations, set practical performance bounds.

This analysis reveals two qualitatively distinct operating regimes. In the small phase shift regime, collective QLS enables efficient parameter estimation over a broad dynamic range that can remain independent of $N_S$, making it well suited to robust, high-throughput clock readout. Increasing the number of logic ions further improves sensitivity, enhances robustness to detection noise, and expands the usable estimation range. In the large phase shift regime, the same mapping approaches a parity-type measurement of the spectroscopy ensemble. In this  limit, parity observables can be accessed using even a single logic ion or without single-ion-resolved detection. This regime can provide enhanced sensitivity near a known operating point, but at the cost of reduced dynamic range, as expected for measurements tailored to highly sensitive nonclassical states.

Our results establish collective QLS as a versatile and scalable interface for many-ion spectroscopy, tunable between robust clock readout and high-sensitivity measurements of nonclassical observables.

\section{Collective QLS protocol}
We begin with a theoretical analysis and generalization of the collective QLS protocol recently demonstrated in Ref.~\cite{cui2022scalable}. The protocol, sketched in Fig.~\ref{fig:qls_sequence}, uses the shared motional mode as a quantum bus to transfer information about the spectroscopy ensemble onto a logic ensemble that can be read out with high fidelity. The protocol consists of four steps: (i) the logic ions are entangled with the mode through a spin-dependent displacement $\hat U_{\mathrm L}$; (ii) the spectroscopy ions imprint their ground-state population onto the mode via a second displacement $\hat U_{\mathrm S}$; (iii) the initial logic–-motion entangling operation is reversed by $\hat U_{\mathrm L}^\dagger$; and (iv) a projective measurement $M_{\mathrm L}$ on the logic ions yields an estimate of the spectroscopy ground-state population prior to the QLS sequence. Because the population imbalance after Ramsey or Rabi interrogation encodes the interferometric parameter, this readout directly enables clock-relevant parameter estimation using measurements on the logic ions alone.

To describe the protocol we introduce the collective spin operators of the spectroscopy and logic ensembles,
$\hat S_\alpha=\frac12\sum_{k}\hat\sigma_\alpha^{k,S},\ \hat L_{\alpha}=\frac12\sum_{k}\hat \sigma_\alpha^{k,L}$  (with $\alpha=x,y,z)$, where $\hat \sigma_\alpha^{k,L}$ are Pauli matrices acting on the states $\ket{\downarrow},\ket{\uparrow}$ of ion $k$ of the $K=S,L$ spectroscopy or logic ions respectively. The shared motional-mode is described by bosonic creation and annihilation operators $\hat a^\dagger$ and $\hat a$.

The sequence starts from a product state of the logic ions polarized along the $x$-axis, $\ket{\psi_L}\propto(\ket{\downarrow}+ \ket{\uparrow})^{\otimes N_L}$ and with the mode cooled to the motional ground state $\ket{0}$. The spectroscopy ions are prepared in an arbitrary state $\ket{\psi_S}$. Similar to direct fluorescence measurements, our goal is to determine the ground-state population, $\braket{S_\downarrow} = N_S/2-\braket{S_z}$, via measurements on the logic ensemble.  Since we know $N_S$ exactly, a measurement of $\braket{S_z}$ is tantamount to a measurement of $\braket{S_\downarrow}$.

In step (i), the logic ions are coupled to the mode via a spin-dependent displacement described by
\begin{align}
    \hat H_L=\frac{g_L}{\sqrt{N}}\,(\hat a+\hat a^\dagger)\,\hat L_z,
    \label{eq:H_L}
\end{align}
applied for an interaction time $t_L$, where $N=N_S+N_L$ is the total number of ions. This interaction generates a motional displacement conditioned on the collective spin projection $\hat L_z$, where $g_L$ denotes the logic--motion coupling strength. The required spin-dependent displacement can be generated in several experimentally established ways, for example using light-shift forces or bichromatic fields that drive red and blue motional sideband \cite{molmer1999multiparticle,haljan2005spin,hempel2013entanglement}. These implementations may differ in their technical advantages and limitations, but the theoretical analysis presented here is agnostic to this choice; further discussion on the side band implementation is provided in App.~\ref{app:SpinDependentDisplacement}. In step (ii), the spectroscopy ions are coupled through an orthogonal displacement in phase space,
\begin{align}
\hat H_S=\frac{g_S}{\sqrt{N}}\, i(\hat a-\hat a^\dagger)\,\hat S_{\downarrow},
\label{eq:H_S}
\end{align}
applied for a time $t_S$ and with respective spectroscopy-motion coupling $g_S$. The relative phase between $\hat a$ and $\hat a^\dagger$ makes the two displacements act on orthogonal quadratures, which is essential for mapping the spectroscopy population onto a measurable rotation of the logic ensemble. Furthermore, in this case the displacement on the bosonic mode depends on the number of spectroscopy atoms in the ground-state $\frac{N_S}{2}\left(\mathds{1}-\frac{2\hat S_z}{N_S}\right)$, rather than directly on $\hat S_z$, where $\mathds{1}$ is the identity operator. We consider this form of the spectroscopy-state-dependent displacement, which is implemented in $\mathrm{Al}^+$ optical clocks by coupling the ground state to an auxiliary third level~\cite{cui2022scalable}. We note, however, that the following discussion applies to couplings proportional to $S_z$, for example by performing the measurement in a rotated basis. While the qualitative considerations remain the same, the effects of noise and higher-order corrections can depend on the specific form of the coupling. Finally, step (iii) implements an echo of the logic ion dependent displacement by applying $-\hat H_L$ for the same duration $t_L$, which disentangles the logic ions from the bosonic mode. 

Using the braiding relations of bosonic displacement operators, the joint state after the sequence, acting on any initial state
$\ket{\psi_0}=\ket{\psi_L}\otimes\ket{\psi_S}\otimes\ket{0}$,
can be written as
\begin{align}
\ket{\psi}
&= e^{+i \hat H_L t_L}\,e^{-i \hat H_S t_S}\,e^{-i \hat H_L t_L}\;
\ket{\psi_0}\notag\\
&= e^{-i \hat H_S t_S}\, e^{-i\theta\,\hat S_{\downarrow} \hat L_z}\ket{\psi_0},
\label{eq:post_qls_state}
\end{align}

At the end of the protocol, the logic ensemble is again decoupled from the motional mode. During the sequence, however, the logic ensemble acquires a geometric phase that effectively implements a controlled rotation 
\begin{align}
    \hat U(\theta) = e^{-i\theta\,\hat S_{\downarrow} \hat L_z}
    \label{eq:U_SL}
\end{align}
about the $z$-axis by an angle that depends on the state of the spectroscopy ions, $\theta\,\hat S_{\downarrow}$, with
\begin{align}
    \theta=\frac{2 g_L t_L\, g_S t_S}{N}.
    \label{eq:information_transfer}
\end{align}
We refer to $\theta$ as the interaction parameter of the protocol: increasing $\theta$ amplifies the spectroscopy-dependent rotation imprinted on the logic ensemble.

Since the residual spectroscopy--motion interaction, $e^{-i \hat H_S t_S}$ is independent of the logic degrees of freedom, it does not affect the statistics of measurements performed on the logic ensemble. It can therefore be ignored when considering observables that act only on the logic ions. As a result, the expectation value of an appropriate logic observable, such as $\langle \hat L_{x,y}\rangle$, encodes the spectroscopy ground-state population.

The reduced density matrix of the logic ensemble, after tracing out the mode and the spectroscopy subsystem in Eq.~\eqref{eq:post_qls_state}, becomes
\begin{align}
\hat \rho_{L}
&=\sum_{n_{\downarrow}=0}^{N_S} {n_{\downarrow}}\,
e^{-i\theta {n_{\downarrow}} \hat L_z}\,\ket{\psi_L}\bra{\psi_L}\,e^{+i\theta {n_{\downarrow}} \hat L_z},
\label{eq:post_qls_dm}
\end{align}
where ${n_{\downarrow}} = |\braket{n_{\downarrow}|\psi_S}|^2$ is the probability to find $n$ spectroscopy ions in the ground state. Here $\ket{n_{\downarrow}}$ denotes an eigenstate of $\hat S_z$, defined by
$\hat S_{\downarrow}\ket{n_{\downarrow}}=n_{\downarrow}\,\ket{n_{\downarrow}}$ for $ n_{\downarrow}\in\{0,1,\dots,N_S\}.$
Equation~\eqref{eq:post_qls_dm} shows that, from the perspective of the logic ions, the QLS sequence implements a classical mixture of $z$-rotations by angles $\theta {n_{\downarrow}}$, i.e., a dephasing channel whose strength is set by the distribution $\{{n_{\downarrow}}\}$ of the spectroscopy $\hat S_z$ population. In the special case that the spectroscopy ensemble is prepared in an $\hat S_z$ eigenstate, the mixture collapses to a single unitary: the logic ensemble disentangles from both the spectroscopy ions and the mode, and the net effect of the QLS sequence reduces to a coherent rotation of the logic spins \cite{gilmore2021quantum}.

\section{Metrological performance}
\label{sec:metrological_performance}
\begin{figure}[t]
    \centering
    \includegraphics[]{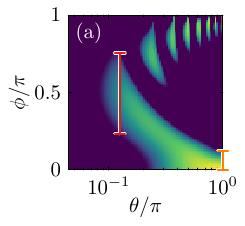}
    \includegraphics[]{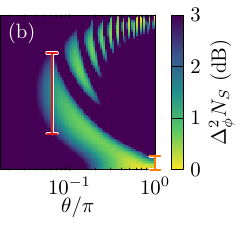}
\caption{
    \textbf{Dependence of the sensitivity on the interaction parameter.}
    The panels show the sensitivity, quantified by the estimator variance $\Delta_{\phi}^2$, for estimating the spectroscopy-state parameter $\phi$ from measurements of the logic ions after the QLS sequence. The sensitivity is shown as a function of the interaction parameter $\theta$ and $\phi$. Panels (a) and (b) correspond to $N_S=N_L=8$ and $N_S=N_L=16$, respectively. Values with $\Delta_{\phi}^2 \geq 2/N_S$ are shown in purple, while smaller variances are color-coded according to the color bar. The red and orange lines indicate $\theta=\pi/N_S$ and $\theta=\pi$, respectively, together with the corresponding dynamic ranges. For $\theta=\pi/N_S$, the dynamic range remains unchanged as the number of ions increases, whereas for $\theta=\pi$ it decreases with increasing ion number.
}
    \label{fig:Sensitivty}
\end{figure}

We now analyze the parameter-estimation performance of collective QLS for a spectroscopy spin-coherent state, as depicted in Fig.~\ref{fig:qls_sequence}, where the parameter $\phi$ is encoded in the ground--excited population imbalance of the spectroscopy ions. Concretely, we consider the product state
\begin{align}
    \ket{\psi_{S,\phi}}
    =
    \left(\cos(\phi/2)\ket{\uparrow}+\sin(\phi/2)\ket{\downarrow}\right)^{\otimes N_S},
    \label{eq:Spectroscopy_CSS}
\end{align}
as generated by a Ramsey (or equivalently, Rabi) interrogation sequence.

Having established how the QLS sequence maps the spectroscopy population distribution onto the logic ensemble, we now quantify how well this mapping can be used for parameter estimation. After the QLS sequence, the information about the spectroscopy population is extracted from measurements on the logic ions alone. We focus on the collective logic observable $\langle \hat L_x\rangle$ and estimate $\phi$ from the sample mean obtained from independent experimental repetitions. 

At first sight, preparing the logic ensemble along the $x$ axis and subsequently measuring in the same basis may appear counterintuitive. However, we consider two distinct operational regimes in which an $x$-basis measurement is naturally suited to the resulting logic states. In the first regime, the logic ensemble is rotated on average by approximately $\pi/2$. In the second, it is either left unrotated or rotated by approximately $\pi$, producing two states that are optimally distinguished through a measurement of $\hat L_x$.

Using standard error propagation, the corresponding estimator variance is
\begin{align}
\Delta^2_{\phi}
=
\frac{\Delta^2_{L_x}}{\left(\frac{\partial}{\partial\phi}\braket{\hat L_x}\right)^2},
\label{eq:estimator_variance}
\end{align}
where $\Delta^2_{L_x}=\braket{\hat L_x^2}-\braket{\hat L_x}^2$. Closed-form expressions for the logic expectation value $\braket{\hat L_x}$ and variance $\Delta^2_{L_x}$ are given in App.\ref{app:mean_values}. In addition to the local sensitivity captured by Eq.~\eqref{eq:estimator_variance}, we characterize the usable parameter interval by defining the dynamic range $D_\phi$ as the range of $\phi$ over which the estimator variance remains below twice the spectroscopy SQL, $\Delta_\phi^2\le 2/N_S$.

Figure~\ref{fig:Sensitivty} summarizes the dependence of the estimator variance on the interaction parameter $\theta$ and the operating point $\phi$. For very small $\theta$, the spectroscopy-dependent rotation imprinted on the logic ensemble is too weak to resolve the population imbalance. Useful readout emerges only once $\theta$ reaches a scale of order $\pi/N_S$, because only then  the accumulated rotation is large enough to  distinguish   different spectroscopy populations. In this small-interaction regime, the best sensitivity occurs near $\phi=\pi/2$, where the population imbalance is most sensitive to the encoded parameter. As $\theta$ increases, the optimal operating point continuously shifts toward $\phi=0$, eventually approaching a parity-sensitive regime. This improves the local sensitivity near a known operating point but simultaneously narrows the region of $\phi$ over which the sensitivity remains close to optimal. The following analysis makes this tradeoff between sensitivity and dynamic range explicit.

\begin{figure}[t]
    \centering
    \includegraphics[]{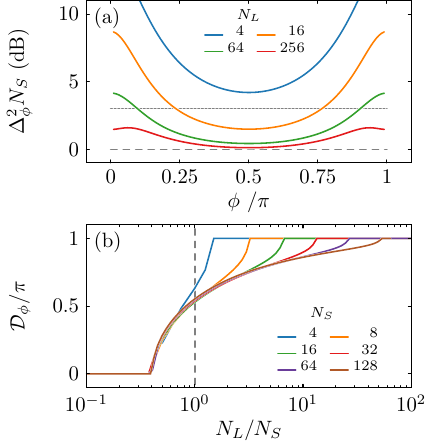}
    \caption{
    \textbf{Sensitivity and dynamic range in the small-interaction regime.}
    (a) Estimator variance, $\Delta_{\phi}^2$, plotted as a function of the parameter $\phi$ for fixed interaction $\theta=\pi/N_S$ and $N_S=16$ spectroscopy ions. Colors indicate the number of logic ions $N_L$. The horizontal dashed and dotted  lines indicate $\Delta_{\phi}^2 =1/N_S$ and $\Delta_{\phi}^2 =2/N_S$, respectively. 
    (b) Dynamic range $\mathcal{D}_{\phi}$, defined as the range over which $\Delta{\phi}^2 \leq 2/N_S$, as a function of the ratio between logic and spectroscopy ions, $N_L/N_S$, for $\theta=\pi/N_S$. Colors indicate different numbers of spectroscopy ions, and the vertical dashed line indicates $N_L=N_S$.
}
    \label{fig:dynamic_range}
\end{figure}

\paragraph{Small-interaction, $\theta=\mathcal{O}(1/N_S)$.}

As highlighted in Fig.~\ref{fig:Sensitivty}, in the small-interaction regime the largest sensitivity is obtained near $\phi=\pi/2$. At this operating point the spectroscopy state has support over the full range of $\hat S_\downarrow$ eigenstates, so the QLS readout must distinguish many possible spectroscopy ground-state populations. For a fixed eigenstate with $n_\downarrow$ atoms in the ground state, the QLS sequence produces a coherent rotation of the logic ensemble by an angle $\theta n_\downarrow$. Thus, resolving the spectroscopy population distribution requires the set of logic states associated with different values of $n_\downarrow$ to be sufficiently distinguishable.

This gives a simple interpretation of the optimal small-interaction scale. If $\theta$ is too small, all possible spectroscopy populations produce nearly the same logic state and little information is transferred. By contrast, choosing
$\theta=\pi/N_S$ spreads the logic states associated with $n_\downarrow=0,\ldots,N_S$ across half of the equator of the collective Bloch sphere: the $n_\downarrow=0$ component leaves the logic spin pointing along the $x$ axis, while the $n_\downarrow=N_S$ component rotates it to the opposite direction. This choice therefore provides an efficient compromise between distinguishability of different spectroscopy populations and maintaining a broad, rotation response. At this operating point, one obtains the asymptotic expression
\begin{align}
    \lim_{\substack{\phi\rightarrow\pi/2\\\theta\rightarrow\pi/N_S}} \Delta^2_\phi
    =\frac{1}{N_S}\left(1+\frac{4}{\pi^2}\frac{N_S}{N_L}\right)
    +\mathcal O\!\left(1/{N_S^2}\right).
    \label{eq:smalltheta_scaling}
\end{align}
The first term is the spectroscopy projection-noise contribution, while the second term arises from projection noise of the logic ensemble. The latter can be suppressed by increasing the number of logic ions. Therefore, retaining SQL scaling, $\Delta_\phi^2\sim 1/N_S$, in this small-interaction regime requires the number of logic ions to scale proportionally with the number of spectroscopy ions, $N_L=\mathcal{O}(N_S)$.

Beyond the minimum variance, practical sensing also requires a sufficiently large \emph{dynamic range}, i.e., a broad interval of $\phi$ over which the estimator variance remains close to its optimum. This behavior is illustrated in Fig.~\ref{fig:dynamic_range}. Panel (a) shows that, at fixed $\theta=\pi/N_S$, increasing $N_L$ lowers the estimator variance near $\phi=\pi/2$ and broadens the region over which near-SQL sensitivity is maintained. Panel (b) makes this scaling explicit: for small logic ensembles the useful parameter window is narrow, whereas for $N_L$ comparable to or larger than $N_S$ the dynamic range approaches an $N_S$-independent value. Thus, Fig. \ref{fig:dynamic_range} confirms that scaling the logic ensemble together with the spectroscopy ensemble preserves not only the SQL sensitivity predicted by Eq.~\eqref{eq:smalltheta_scaling}, but also a finite usable estimation range.

\paragraph{Large-interaction, $\theta = \mathcal{O}(1)$.}

\begin{figure}[t]
    \centering
    \includegraphics[]{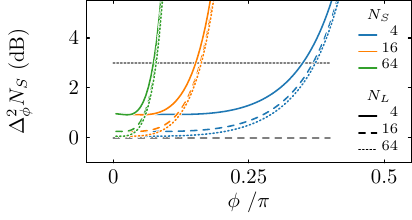}
    \caption{
    \textbf{Sensitivity and dynamic range in the large-interaction regime.}
    Estimator variance, $\Delta^2_{\phi}$, as a function of the parameter $\phi$ at fixed interaction parameter $\theta=\pi/2$. Colors indicate the spectroscopy-ion number, while line styles indicate the logic-ion number. The horizontal dashed and dotted lines indicate $\Delta_{\phi}^2 =1/N_S$ and $\Delta_{\phi}^2 =2/N_S$, respectively.}
    \label{fig:large_theta}
\end{figure}

We now turn to the large-interaction regime, where $\theta$ remains finite as the spectroscopy ensemble grows. Because the unitary in Eq.\eqref{eq:U_SL} is periodic in $\theta$, the largest useful value is $\theta=\pi$; increasing $\theta$ beyond this point does not provide additional information.

In this regime, the optimal operating point shifts toward $\phi=0$. Near this point, the spectroscopy ensemble predominantly occupies a superposition of the fully excited state and the inverted W state, i.e., the permutation-symmetric state with a single atom in the ground state and all remaining atoms excited. The QLS sequence then leaves the logic ensemble in a classical mixture of two coherent spin states: its initial state and the same state rotated by an angle $\theta$ about the $z$ axis. The weight of the rotated component is set by the population of the spectroscopy ensemble in the inverted W state \cite{dur2000three}. In the limit $\phi\to0$, the estimator variance becomes
\begin{align}
\lim_{\phi\rightarrow 0} \Delta^2_{\phi}
=
\frac{1}{N_S}
\left(
1+\frac{1}{N_L}\cot^2\frac{\theta}{2}
\right).
\label{eq:large_transfer_phi0}
\end{align}
The leading contribution is again the spectroscopy SQL. The additional term arises from projection noise of the logic ensemble, but, in contrast to the small-interaction regime, it is suppressed by $N_L$ alone and does not require $N_L$ to scale with $N_S$. 
Thus, for fixed $\theta$ and fixed $N_L$, the sensitivity at $\phi\to0$, normalized to the spectroscopy SQL, is independent of the number of spectroscopy ions.
In the special case $\theta=\pi$, the logic-noise contribution vanishes entirely, and SQL sensitivity is obtained even with a single logic ion.

Equation\eqref{eq:large_transfer_phi0} also contains information on  how large $\theta$ must be to retain SQL scaling near $\phi=0$. The crossover occurs when $\frac{1}{N_L}\cot^2\frac{\theta}{2}\sim 1$,
which for small $\theta$ gives $\theta\sim 2/\sqrt{N_L}$. At this crossover scale, setting $\theta=2/\sqrt{N_L}-2/\left(3 N_L^{3/2}\right)$ yields
\begin{align}
\lim_{\phi\rightarrow 0} \Delta_{\phi}^2
=
\frac{2}{N_S}
+\mathcal{O}\left(1/N_L^2\right).
\end{align}
Hence, increasing the number of logic ions allows one to reach SQL-scaling sensitivity at smaller interaction strength.

The advantage of the large-interaction regime is therefore a favorable local sensitivity near $\phi=0$, even for a fixed number of logic ions. However, this comes at the cost of a reduced dynamic range. At $\theta=\pi$, the estimator variance for generic $\phi$ is
\begin{align}
\lim_{\theta\rightarrow \pi} \Delta^2_{\phi}
= \frac{\cos^{2-2N_S}(\phi)-\cos^2(\phi)}
{N_S^2\sin^2(\phi)},
\label{eq:large_theta_scaling}
\end{align}
which is minimized at $\phi=0$ with value $1/N_S$. Expanding away from $\phi=0$ shows that the parameter value at which the variance doubles scales as $\phi=\mathcal{O}(1/\sqrt{N_S})$. Thus, although the variance at the optimal point retains SQL scaling, the interval over which this sensitivity is maintained shrinks as the spectroscopy ensemble grows.

This behavior is illustrated in Fig.~\ref{fig:large_theta}. The curves for different $N_S$ collapse near $\phi=0$ when the variance is normalized to the SQL, confirming that the small-$\phi$ sensitivity relative to $1/N_S$ is independent of $N_S$ for fixed $N_L$. At the same time, the low-variance region becomes progressively narrower with increasing $N_S$, demonstrating the loss of dynamic range. Fig.~\ref{fig:large_theta} therefore makes explicit the central tradeoff of the large-interaction regime: one can achieve sensitivities scaling as the SQL as $1/N_S$ near a known operating point without scaling the number of logic ions, but the usable parameter window decreases with spectroscopy-ensemble size.

\section{QLS parity measurement}
\label{sect:ParityMeasurement}
\begin{figure}[t]
    \centering
    \includegraphics[]{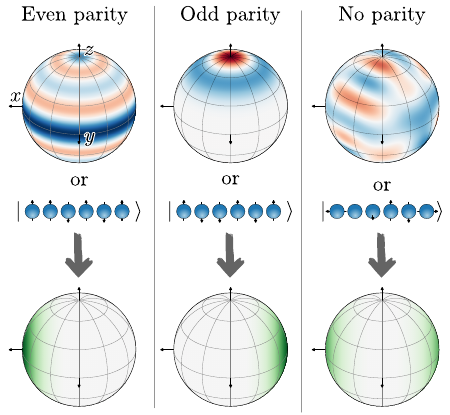}
    \caption{
    \textbf{Parity mapping through the QLS sequence}
    Even-, odd-, and no-parity states of the spectroscopy ions are mapped through the quantum-logic sequence onto coherent spin states of the logic ions pointing along the $+x$-axis, the $-x$-axis, or to a statistical mixture of both, respectively. The relative weight of the two orientations reflects the projection of the initial state onto the corresponding parity subspaces. The top row schematically illustrates entangled, permutation-symmetric states restricted to the even, odd and no-parity subspace, while the middle row shows examples of unentangled product states. Regardless of the specific spectroscopy state, all are mapped onto the logic-ion spin states depicted in the bottom row.}
    \label{fig:parity_measurement}

\end{figure}

At the special interaction parameter $\theta=\pi$, the QLS map acquires a particularly simple interpretation: it maps the $z$-basis parity of the spectroscopy ensemble onto the orientation of the logic spin. Parity measurements provide a natural readout for highly sensitive states and can saturate the optimal phase sensitivity for important classes of probes, including GHZ states \cite{bollinger1996optimal, seshadreesan2013phase}. Concretely, the logic ensemble ends in a classical mixture of two macroscopically distinct states: the initial state polarized along the $x$-axis and the state polarized in the opposite direction. Which of the two occurs is determined by whether the spectroscopy ensemble lies in the $+$ or $-$ eigenspace of the $z$-basis parity operator $\hat \Pi_z=\prod_{k\in S}\hat \sigma_z^{(k)}$. Formally, a measurement of $\hat L_x$ on the logic ensemble yields
\begin{align}
    \lim_{\theta \to \pi} \braket{\hat L_x }
    &= \frac{N_L}{2}\sum_{n_{\downarrow}=0}^{N_S} p_{n_{\downarrow}}\,(-1)^{n_{\downarrow}}
    \notag\\
    &= \frac{N_L}{2}(-1)^{N_S}\,\braket{\psi_{S}|\hat \Pi_z |\psi_{S}}. 
\label{eq:parity_measurement}
\end{align}
Importantly, Eq.~\eqref{eq:parity_measurement} is completely general: it holds for arbitrary many-body states of the spectroscopy ensemble and does not rely on permutation symmetry.

A key advantage of QLS-based parity detection, beyond enabling the readout of spectroscopy ions that cannot be measured directly, is that it can substantially relax detection-fidelity requirements. This is particularly important for entanglement-enhanced interferometry with GHZ-type probe states. In such protocols, parity measurements provide access to a Heisenberg-limited signal, characterized by an estimator variance scaling as $1/N_S^2$, but are otherwise extremely sensitive to detection errors. Other approaches to mitigating measurement noise similarly rely on converting fragile many-body information into a more robust measurement basis, for example by reversing the unitary used to prepare the entangled probe state \cite{davis2016approaching, frowis2016detecting,anders2018phase,haine2018using, mirkhalaf2018robustifying} or, more generally, by applying a decoding operation before detection. 

Collective QLS realizes a related idea at the level of the readout interface: the spectroscopy parity is mapped onto the logic ensemble, which is left in a statistical mixture of two coherent spin states polarized along opposite directions of the $x$ axis. The measurement therefore reduces to distinguishing two macroscopically separated logic states, rather than reconstructing a many-body parity from individual spectroscopy-ion detection outcomes. In particular, it requires neither individual addressing of the spectroscopy ions nor site-resolved fluorescence detection. As a result, detection errors affect the inferred parity only through the collective logic readout channel and are far less severe than in direct spectroscopy-parity detection.

In the following, we model imperfect readout by assuming independent detection errors, such that each ion’s measurement outcome is flipped with probability $\varepsilon$. We focus in particular on the case where the spectroscopy ensemble is prepared in the GHZ state in the $x$-basis, 
$\ket{\psi_{\rm GHZ}}
=
e^{-i\frac{\pi}{2}\hat S_y}
\frac{\ket{\uparrow}^{\otimes N_S}+\ket{\downarrow}^{\otimes N_S}}{\sqrt{2}}$.
In the absence of detection noise, a parity measurement at the Heisenberg limit . In the presence of imperfect readout, however, preserving this Heisenberg scaling requires the single-ion misidentification probability to decrease with system size as $\varepsilon \propto 1/N_S$ (see App.~\ref{app:direct_parity}). This increasingly stringent requirement on detection fidelity constitutes a major practical obstacle to observing entanglement-enhanced sensitivities with large GHZ states.

By contrast, for the QLS-based parity readout, the detection-noise penalty does not increase exponentially with $N_S$ and is further suppressed by a factor of $1/N_L$ as discussed in App.~\ref{app:QLS_parity}. As a result, Heisenberg-limited parameter estimation no longer requires $\varepsilon$ to decrease with $N_S$. Instead, for fixed $\varepsilon$, the remaining penalty can be mitigated by increasing the size of the logic ensemble.

\section{Experimental requirements}

So far, we have focused on the fundamental precision limits of the collective QLS mapping in the idealized setting. In a realistic implementation, however, the protocol will inevitably be affected by additional noise sources and experimental imperfections. In the following, we examine several of these corrections and determine the level of control required for collective QLS to remain scalable under such nonideal conditions.

\subsection{Protocol duration}
A necessary  requirement is  that the collective QLS sequence be executed on a timescale short compared to the relevant coherence times of the dominant error mechanisms in the trap.

The total duration of the QLS sequence is $t=t_S+2t_L$, where $t_L$ and $t_S$ are the interaction times of the logic- and spectroscopy-ion displacements, respectively. In practice, however, the QLS interaction may constitute only a small fraction of the experimental cycle, since laser cooling, state preparation, and fluorescence detection in state-of-the-art optical ion clocks typically require several milliseconds and can introduce total dead times of order $10\,\mathrm{ms}$. For fixed interaction parameter in Eq.~\eqref{eq:information_transfer} one can minimize $t$ by optimizing the partitioning of the interaction time between the two spin dependent displacements. This yields $t^{*}_S = \sqrt{\frac{N\theta}{g_S g_L}}$, and $t_L^* = \frac{1}{2}\,t_S^*,$ for the optimal times 
and thus a minimum total duration is $t_S^* + 2t_L^* = 2\sqrt{\frac{N\theta}{g_L g_S}}.$ 

It is instructive to evaluate $t^*=t_S^*+2t_L^*$ in the two operating regimes identified in Sec.~\ref{sec:metrological_performance}. In doing so, we adopt the scaling assumption introduced in Eqs.~\eqref{eq:H_L} and \eqref{eq:H_S}: the single-ion Lamb--Dicke parameter is held fixed as the system size increases, such that the Lamb--Dicke parameter associated with the collective crystal mode scales as $\eta\propto 1/\sqrt{N}$.In the large-interaction regime, $\theta=\mathcal{O}(1)$, one obtains $\sqrt{g_L g_S}t^*\propto \sqrt{N}$, i.e. the protocol time grows with the square root of the total ion number. In contrast, in the small-interaction regime, $\theta=\mathcal{O}(1/N_S)$, the minimum duration scales as $\sqrt{g_L g_S}t^*\propto \sqrt{N/N_S}$. 
As discussed in Sec.~\ref{sec:metrological_performance}, retaining SQL-like scaling as $N_S$ increases requires $N_L\propto N_S$. Under this condition, $N/N_S=\mathcal{O}(1)$, and, assuming a fixed single-ion Lamb--Dicke parameter, the optimized protocol duration remains $\sim 1/\sqrt{g_Lg_S}$, independent of the spectroscopy-ensemble size. This represents a conservative scaling for linear rf Paul traps: if the axial confinement is reduced as the chain length increases to remain below the linear-to-zigzag instability \cite{morigi2004dynamics}, the corresponding increase in the single-ion Lamb--Dicke parameters enhances the spin--motion couplings $g_L$ and $g_S$, causing the optimized protocol duration to decrease with ion number.

\begin{table*}[t]
\centering
\renewcommand{\arraystretch}{1.35}
\begin{tabular}{
p{0.1\textwidth}
@{\hspace{2.5em}}
p{0.25\textwidth}
@{\hspace{2.5em}}
p{0.25\textwidth}
@{\hspace{2.5em}}
p{0.25\textwidth}
}
\hline\hline
&
CSS, $\theta=\pi/N_S$, $\phi=\pi/2$
&
CSS, $\theta=\pi$, $\phi\simeq 0$
&
GHZ, $\theta=\pi$, $\phi\simeq \pi/(2N_S)$
\\
\hline

Scaling
&
$\Delta_\phi^2=\mathcal O(1/N_S)$
&
$\Delta_\phi^2=\mathcal O(1/N_S)$
&
$\Delta_\phi^2=\mathcal O(1/N_S^2)$
\\

\hline

Mode-frequency fluctuations
&
From Eq.~\eqref{eq:detuning_small_transfer}: requires $N_L=\mathcal O(N_S)$ and 
$
\frac{\sigma_\delta^2}{g_S^2}\left(1+2\bar{n}\right)
=
\mathcal O(1).
$
&
From Eq.~\eqref{eq:detuning_large_transfer_css}: requires 
$
\phi\sim1/\sqrt{N_S}
$
and
$
\frac{\sigma_\delta^2}{g_S^2}
\left(
1+2\bar n
\right)
=
\mathcal O(1/N_S).
$
&
From Eq.~\eqref{eq:detuning_large_transfer_ghz}: requires
$\sigma_\delta^2=\mathcal{O}(N_S^{-2})$ for $N_L=\mathcal{O}(1)$ and $\sigma_\delta^2=\mathcal{O}(N_S^{-3/2})$ for $N_L\sim N_S$.
\\

\hline

Beyond Lamb--Dicke corrections
&
From Eq.~\eqref{eq:LD_small_transfer}: requires $N_L=\mathcal O(N_S)$ and $\bar n\lesssim N_S$ or smaller.
&
From Eq.~\eqref{eq:LD_large_transfer_css}: requires
$
\eta^2 \phi\, \tilde\theta
\left(
1+2\bar n
\right)
\lesssim
N_S
$.
&
From Eq.~\eqref{eq:LD_large_transfer_ghz}: 
and 
$
\eta^2\tilde\theta
=\mathcal{O}(1/N_S^3).$
\\
\hline\hline
\end{tabular}
\caption{
Scaling requirements for retaining SQL or Heisenberg-limited behavior in the presence of mode-frequency fluctuations and leading beyond-Lamb--Dicke corrections. In the last two columns the corrections at $\theta=\pi$ vanish and thus the corrections are evaluated at $\theta=\pi-\tilde{\theta}$, and for all cases $t_S=2t_L$.  
}
\label{tab:imperfection_scaling_requirements}
\end{table*}

\subsection{Noise due to thermal motion}

A practical advantage of the collective QLS protocol, shared by other thermal-motion QLS approaches, is that it does not require ground-state cooling of the common motional mode \cite{tan2015multi,kienzler2020quantum}. Residual finite-temperature effects may nevertheless become relevant in the presence of corrections beyond the Lamb–Dicke approximation \cite{gilmore2021quantum,cui2022scalable}, shifts of the motional-mode frequency arising from intrinsic nonlinearities of the ion crystal \cite{johnson2025nonlinear}, or other imperfections that produce a net detuning between the spin-dependent-displacement drive and the bosonic mode.

The insensitivity to the initial motional state follows directly from the structure of the sequence. In the ideal protocol, the operations are arranged such that no residual entanglement between the logic ions and the mode remains at the end of the sequence. As can be seen from Eq.~\eqref{eq:post_qls_state}, the QLS protocol effectively implements a spectroscopy-dependent rotation of the logic ensemble together with a spectroscopy-dependent displacement of the mode, but no direct coupling between the logic ions and the motion. Since the final readout is performed exclusively on the logic ions, the measurement statistics are, in the ideal case, independent of the motional state present before the sequence, and thus insensitive to a nonzero initial thermal phonon occupation.

This feature also highlights the essential role of the initial logic–motion entanglement generated by the first logic-dependent displacement and removed again by the final inverse displacement. As discussed in App.~\ref{app:initial_logic_motion}, if this initial entangling step were omitted, the final logic displacement would leave the logic ions entangled with the mode. This residual entanglement would reduce the measurement contrast in readout performed solely on the logic ions and moreover, render the protocol significantly more sensitive to the average mode occupation.

\subsection{Mode frequency fluctuations}

A mechanism that can spoil the precise time reversal between the two logic-dependent displacements is a finite detuning between the lasers driving the spin-dependent displacements and the motional mode, for example due to fluctuations of either the laser frequency or the mode frequency itself. Such a detuning leaves residual logic--motion entanglement at the end of the sequence, which manifests as a reduction of the signal contrast in the logic ensemble together with an increase in the effective projection noise \cite{gilmore2021quantum}. Assuming that these fluctuations remain constant over the duration of a single QLS sequence, the resulting corrections to the estimator variance can be derived as a function of the detuning, as shown in App.~\ref{app:mode_detuning}. In particular, we model the detuning fluctuations between different runs of the QLS sequence by a Gaussian random variable with zero mean and variance $\sigma_\delta^2$.

The resulting scaling requirements are summarized in Table~\ref{tab:imperfection_scaling_requirements}. In the small-interaction regime, coherent-spin-state spectroscopy retains SQL scaling provided that $N_L=\mathcal O(N_S)$ and the detuning-induced correction remains perturbative. For fixed coupling ratio and fixed detuning-noise strength, this requires the thermal occupation to remain bounded, $\bar n=\mathcal O(1)$, since the leading thermal contribution is controlled by $\sigma_\delta^2 \bar n/(g_Lg_S)$. This requirement is qualitatively different from the leading beyond-Lamb--Dicke correction discussed below, for which the same small-interaction regime can tolerate $\bar n$ growing linearly with $N_S$.

The requirements are more restrictive in the parity-measurement regime. There, preserving the desired scaling requires the fluctuation-induced contribution to remain below the ideal estimator variance. Consequently, for fixed thermal occupation, the detuning fluctuations must decrease with system size as $\sigma_\delta^2=\mathcal O(1/N_S)$. More generally, the relevant constraint is on the temperature-weighted fluctuation strength, 
$
\frac{\sigma_\delta^2}{g_S^2}(2\bar n+N_S)=\mathcal O(1),
$
up to constants and fixed coupling ratios. This applies both to coherent-spin-state spectroscopy, where one aims to retain SQL scaling, and to GHZ-state spectroscopy, where one aims to retain Heisenberg scaling \cite{bollinger1996optimal, giovannetti2004quantum, pezze2018quantum}.

\subsection{Beyond the Lamb--Dicke approximation}

Throughout this manuscript, we employ the Lamb–Dicke approximation underlying the Hamiltonians in Eqs.\eqref{eq:H_L} and \eqref{eq:H_S}, retaining only terms linear in the Lamb–Dicke parameter $\eta=\Delta k\sqrt{\frac{\hbar}{2m\omega_{\rm m}}}\ll1.$ Here, $\hbar\Delta k$ is the momentum transferred by the Raman beams generating the spin-dependent force, $m$ is the mass of a single ion, and $\omega_{\rm m}$ is the motional-mode frequency. Depending on the trap parameters and the mean phonon occupation of the mode, however, the Lamb–Dicke approximation may not always be sufficiently accurate. In App.\ref{app:LambDickeCorrections}, we therefore derive the next-to-leading-order corrections and analyze their effect on the effective QLS unitary.

The beyond Lamb--Dicke  corrections again generate residual logic--motion entanglement and, in addition, give rise to terms proportional to $\hat L_z^3 \hat S_{\downarrow}$ and $\hat L_z \hat S_{\downarrow}^3$, corresponding to nonlinear interactions between the logic and spectroscopy spins. The resulting requirements are summarized in Table~\ref{tab:imperfection_scaling_requirements}. In the small-interaction CSS regime, the leading correction scales as
$\frac{\eta^2}{N_S}\left(1 + \frac{\bar{n}}{N_S}\right)$
Thus SQL scaling is preserved for constant single-ion Lamb--Dicke parameter $\eta$, as long as the thermal occupation does not grow faster than linearly with system size, $\bar n=\mathcal O(N_S)$. This is in contrast to mode-frequency fluctuations, where for fixed detuning-noise strength the thermal occupation must remain $\mathcal O(1)$.

In the parity-readout regime, the leading order-$\eta^2$ correction vanishes exactly at the ideal interaction point $\theta=\pi$. Away from this point, writing $\theta=\pi-\tilde\theta$, the correction is controlled by the small deviation $\tilde\theta$. For a CSS input state, SQL scaling requires
$
\eta^2\phi\tilde\theta
\left(
1+\frac{\bar n}{N_S}
\right)
=
\mathcal O(N_S^{-1}),
$
while for a GHZ input state, Heisenberg scaling requires operation away from the zeros of $\sin(N_S\phi)$ and
$
\eta^2\tilde\theta
\left(
N_S+\frac{\bar n}{N_S}
\right)
=
\mathcal O(N_S^{-2}).
$
At the ideal parity point, the Lamb--Dicke corrections cancel to second order. Realizing this cancellation, however, requires precise tuning to $\theta=\pi$; here, we therefore examine the residual corrections that arise when $\theta$ deviates slightly from this fine-tuned operating point. 
\subsection{Repeated readout}

The collective QLS sequence also allows, in principle, repeated readout of a single spectroscopy-state preparation. The reason is that the spectroscopy ensemble couples to the logic ions and the shared motional mode only through the operator $\hat S_{\downarrow}$. Therefore, in the spectroscopy $z$ basis, the QLS sequence distinguishes sectors with different ground-state population $n_{\downarrow}$ but does not change the corresponding population distribution.

After a QLS measurement, the logic ions and the motional mode can be reinitialized, for example, by resetting the logic state and recooling the mode. This removes the auxiliary degrees of freedom used for readout, while leaving the spectroscopy state affected only by dephasing between sectors with different $n_{\downarrow}$. Since the population distribution itself is preserved, the QLS sequence can be applied repeatedly to extract additional information about the same population imbalance without further disturbing the signal-encoding degree of freedom \cite{wolf2016non,negnevitsky2018repeated,kienzler2020quantum}. In this sense, repeated measurements effectively increase the total readout resource, playing a role similar to using a larger number of logic ions. This is particularly useful when the measurement uncertainty is dominated by logic-ion projection noise, as occurs in the small-interaction regime, and especially when the number of available logic ions is smaller than the number of spectroscopy ions.

\section{Conclusion and outlook}

We have developed a unified metrological picture of collective quantum logic spectroscopy and identified the operating regimes that determine its performance as a scalable readout interface for many-ion spectroscopy. The interaction parameter $\theta$ controls a continuous crossover between two qualitatively distinct limits. In the small-interaction regime, collective QLS maps population information faithfully onto the logic ensemble, enabling SQL-like sensitivity with a dynamic range that can remain independent of $N_S$ when the number of logic ions is scaled proportionally. In the large-interaction regime, the protocol instead becomes sensitive to the discreteness of the spectroscopy population and, at $\theta=\pi$, realizes an effective parity measurement. This regime can access highly sensitive nonclassical observables, but at the cost of a reduced dynamic range. These results show that collective QLS is not a single fixed readout protocol, but a tunable measurement primitive whose operating point can be adapted to the metrological task.

We also identified the main practical requirements for scalable implementations. In the ideal sequence, the final logic readout is insensitive to the initial thermal occupation of the shared motional mode because residual logic--motion entanglement is canceled by the time-reversal step. Finite mode detuning and beyond-Lamb-Dicke corrections provide the dominant intrinsic limitations when this cancellation is imperfect, setting requirements on mode-frequency stability, thermal occupation, and Lamb-Dicke parameters. In addition, the QLS mapping can be applied repeatedly to the same spectroscopy-state preparation, since it preserves the population distribution over $\hat S_{\downarrow}$ sectors. In the parity-readout regime, collective QLS offers a further advantage over direct parity detection of the spectroscopy ions: detection-noise penalties are suppressed by the collective measurement of the logic ensemble, thereby avoiding the exponentially severe dependence on the spectroscopy-system size that otherwise limits GHZ-based metrology.

Several directions follow naturally from this work. On the metrological side, it will be important to optimize estimator strategies beyond simple error propagation and to analyze collective QLS within full clock protocols including decoherence, laser noise, and technical fluctuations. On the control side, more efficient spin--boson operations or optimized decoding sequences could reduce the protocol duration and improve the interaction efficiency. Finally, beyond optical clocks, the parity-mapping capability suggests applications in quantum information processing: combined with local basis rotations, collective QLS provides access to nonlocal Pauli-string observables, which are central to stabilizer measurements and lattice-surgery protocols in quantum error correction \cite{fowler2012surface, horsman2012surface}. Related nonlocal observables also play an important role in diagnosing topological order \cite{levin2006detecting} and in measurement-based preparation of many-body quantum states \cite{sahay2025classifying, feldmeier2026digital}. These directions highlight collective QLS as a flexible interface between many-body spectroscopy, precision metrology, and quantum information processing.

\section{Acknowledgments}
We thank Mason Marshall and Shawn Geller for useful feedback on the manuscript. This work is  supported by the Vannevar Bush Faculty Fellowship, AFOSR FA9550-24-1-0179, the NSF JILA-PFC PHY-2317149 and NSF QLCI awards OMA-2016244, the U.S. Department of Energy, Office of Science, National Quantum Information Science Research Centers, Quantum Systems Accelerator, Office of Naval Research and  NIST. RK acknowledges funding by the German National Academy of Sciences Leopoldina under grant LPDS 2024-08. 

\appendix
\section{Spin dependent displacement}
\label{app:SpinDependentDisplacement}

Spin-dependent displacements are generated by applying a bichromatic laser field with frequency components
$\omega_{\rm r}=\omega_0-\omega_{\rm m}$ and $\omega_{\rm b}=\omega_0+\omega_{\rm m}$,
which resonantly drive the first red and blue motional sidebands, respectively. Here, $\omega_0$ denotes the qubit-transition frequency and $\omega_{\rm m}$ the frequency of the addressed motional mode, which throughout this work is taken to be the center-of-mass mode.

For a single ion coupled to the addressed motional mode, the light–matter interactions generated by the two tones can be written as \cite{leibfried2003quantum}
\begin{align}
\hat H_{\mathrm{r}}
&=
\frac{\Omega}{2}\eta e^{-\eta^2/2} i
\left[
\hat\sigma_{+}e^{i\phi_{\rm r}}
\frac{L_{\hat a^{\dagger}\hat a}^{(1)}(\eta^2)}
{\hat a^{\dagger}\hat a+1},
\hat a
+\mathrm{h.c.}
\right],\notag\\
\hat H_{\mathrm{b}}
&=
\frac{\Omega}{2} \eta e^{-\eta^2/2} i
\left[
\hat\sigma_{+}e^{i\phi_{\rm b}}
\hat a^\dagger
\frac{L_{\hat a^{\dagger}\hat a}^{(1)}(\eta^2)}
{\hat a^{\dagger}\hat a+1}
+\mathrm{h.c.}
\right].
\end{align}
Here, $\Omega$ is the carrier Rabi frequency, $\phi_{\rm r}$ and $\phi_{\rm b}$ are the phases of the red- and blue-sideband tones, and $\eta
=
\Delta k
\sqrt{\frac{\hbar}{2m\omega_{\rm m}}}
$
is the mode-specific Lamb–Dicke parameter for the ion. In this expression, $\hbar\Delta k$ is the momentum transferred by the Raman beams, $m$ is the mass of a single ion, $\omega_{\rm m}$ is the frequency of the addressed motional mode. Finally, $L_{\hat a^\dagger\hat a}^{(1)}(\eta^2)$ denotes an associated Laguerre polynomial evaluated at $\eta^2$, with the phonon-number operator $\hat a^\dagger\hat a$ replacing its degree.

In the Lamb--Dicke regime $\eta\ll 1$, the number-dependent factor can be expanded in powers of $\eta$ as
\begin{align}
f_{\hat a^{\dagger}\hat a}(\eta^2)&\equiv e^{-\eta^2 /2}\,\frac{L_{\hat a^{\dagger}\hat a}^{(1)}(\eta^2 )}{{\hat a^{\dagger}\hat a}+1}\notag \\
&=1-\frac{{\hat a^{\dagger}\hat a}+1}{2}\,\eta^2+\mathcal O(\eta^4).
\end{align}
Truncating this expansion at zeroth order, i.e. setting
$e^{-\eta^2/2}\,L_{{\hat a^{\dagger}\hat a}}^{(1)}(\eta^2)/({\hat a^{\dagger}\hat a}+1)\to 1$,
corresponds to the usual Lamb--Dicke approximation, in which the sideband couplings become independent of the motional occupation.

For an ion crystal of $N$ ions, the coupling to the axial center-of-mass mode is reduced because the mode’s effective mass scales with $N$. Consequently, the single-ion Lamb--Dicke parameter $\eta$ appearing above is replaced by the COM-mode value $\eta_{\rm COM} =\eta/\sqrt{N}$.
Taking this into account, and choosing equal optical phases $\phi_{\rm r}=\phi_{\rm b}=3\pi/2$, the red- and blue-sideband interactions combine to 
\begin{align}
\hat H_{\rm r}+\hat H_{\rm b}\simeq
\frac{g}{\sqrt{N}}\sum_{k=1}\frac{\hat \sigma^{(k)}_x}{2}\left(f_{\hat a^{\dagger}\hat a}\left(\frac{\eta^2}{{N}}\right)\hat a + \mathrm{h.c.}\right),
\end{align}
where  we introduce the spin--motion coupling strength $g=\Omega\,\eta$.

By applying the bichromatic red- and blue-sideband tones to the logic ions, and preceding and following the interaction with a $\pi/2$ rotation about the $y$-axis and its inverse, one obtains the effective Hamiltonian in Eq.~\eqref{eq:H_L}. The overall sign of this Hamiltonian can be reversed by changing the optical phases to $\phi_{\rm r}=\phi_{\rm b}=\pi/2$.

To realize an interaction of the form in Eq.~\eqref{eq:H_S}, we instead choose
$\phi_{\rm r}=0$ and $\phi_{\rm b}=\pi$ and drive a sideband transition between the qubit ground state
and an auxiliary internal level. As for the logic interaction, the desired spin operator is obtained by applying the corresponding $\pi/2$ pulses on this transition before and after the bichromatic drive.

\section{Logic-spin moments after the QLS sequence}
\label{app:mean_values}

Here we derive the expectation value and variance of the collective logic observable ${\hat L}_x$ after the QLS sequence. Starting from Eq.\eqref{eq:post_qls_dm}, the logic ensemble is a classical mixture of coherent spin states rotated about the $z$ axis by angles $\theta n_\downarrow$, where $n_\downarrow$ is the number of spectroscopy ions in the ground state. Denoting the corresponding probability distribution by $p_{n_\downarrow}$, one obtains
\begin{align}
\braket{{\hat L}_x}
=
\frac{N_L}{2}
\sum_{n_\downarrow=0}^{N_S}
p_{n_\downarrow}
\cos(\theta n_\downarrow),
\label{eq:Jx_general}
\end{align}
where we used that the initial logic state is polarized along $x$. Similarly,
\begin{align}
\braket{\hat{L}_x^2}
=&
\frac{N_L^2}{4}
\sum_{n_\downarrow=0}^{N_S}
p_{n_\downarrow}
\cos^2(\theta n_\downarrow)\notag \\
&+
\frac{N_L}{4}
\sum_{n_\downarrow=0}^{N_S}
p_{n_\downarrow}
\sin^2(\theta n_\downarrow).
\label{eq:Jx2_general}
\end{align}
These expressions hold for an arbitrary spectroscopy input state, with $p_{n_\downarrow}$ denoting its population distribution in the $S_\downarrow$ basis.

For the spectroscopy spin-coherent state considered in Sec.\ref{sec:metrological_performance},
\begin{align}
p_{n_\downarrow}
=
\binom{N_S}{n_\downarrow}
\sin^{2n_\downarrow}\left(\frac{\phi}{2}\right)
\cos^{2(N_S-n_\downarrow)}\left(\frac{\phi}{2}\right),
\end{align}
the sums can be evaluated in closed form. Defining $
z
=
\cos\left(\frac{\theta}{2}\right)
-
i\sin\left(\frac{\theta}{2}\right)\cos\phi,
 \tilde z
=
\cos\theta
-
i\sin\theta\cos\phi,
$ we find
\begin{align}
\braket{{\hat L}_x}
=
\frac{N_L}{2}
\Re\left[
e^{i\theta N_S/2}z^{N_S}
\right],
\label{eq:Jx_CSS}
\end{align}
and
\begin{align}
\braket{{\hat L}_x^2}
=
\frac{N_L}{8}
\left[
N_L+1
+
(N_L-1)
\Re\left(
e^{i\theta N_S}\tilde z^{N_S}
\right)
\right].
\label{eq:Jx2_CSS}
\end{align}
Differentiating Eq.\eqref{eq:Jx_CSS} with respect to $\phi$ gives
\begin{align}
\frac{\partial}{\partial\phi}\braket{{\hat L}_x}
=
-&\frac{N_S N_L}{2}
\sin\left(\frac{\theta}{2}\right)
\sin\phi\notag \\
&\cdot\Im\left[
e^{i\theta N_S/2}z^{N_S-1}
\right].
\label{eq:dJx_CSS}
\end{align}
Equations\eqref{eq:Jx_CSS}--\eqref{eq:dJx_CSS} provide the ingredients needed to evaluate the estimator variance in Eq.~\eqref{eq:estimator_variance}.

\section{Detection noise}
\label{app:DetectionNoise}

Here we analyze the effect of detection errors on the parity readout discussed in Sec.\ref{sect:ParityMeasurement}. Measurements in the $x-y$ plane can be implemented by a final basis rotation followed by $z$-basis detection, so it is sufficient to consider a generic $z$-basis observable $\mathcal M_z$ subject to independent spin-flip errors. A single-site error on spin $k$ acts as
\begin{align}
\hat \Lambda_k(\mathcal M_z)
=
(1-\varepsilon)\hat{\mathcal M}_z
+
\varepsilon \hat \sigma_k^x \hat{\mathcal M}_z\hat \sigma_k^x,
\end{align}
and the full detection channel for $N$ measured spins is $\hat \Lambda
=
\hat \Lambda_1\circ\cdots\circ\hat \Lambda_N$.
Noisy expectation values are evaluated in the Heisenberg picture as
$\braket{\hat {\mathcal M}_z}_{\varepsilon}
=
\tr\left[
\hat \rho_{\phi}\Lambda(\hat{\mathcal M}_z)
\right],$
where $\hat \rho_{\phi}$ denotes the state after any pre-readout basis rotations.

\subsection{Direct parity readout on spectroscopy ions}
\label{app:direct_parity}

For a direct measurement of the spectroscopy parity $\hat\Pi_z$, each detection error flips the sign of the measured parity. Hence only the parity of the number of detection errors matters, giving
\begin{align}
\hat \Lambda(\hat \Pi_z)
=
(1-2\varepsilon)^{N_S}\hat \Pi_z,
\qquad
\hat \Lambda(\hat \Pi_z^2)
=
\Pi_z^2
=
\mathds{1}.
\end{align}
Estimating $\phi$ from the sample mean of $\hat \Pi_z$ over repeated measurements yields
\begin{align}
\Delta^2_{\hat \Pi_z,\varepsilon}
&=
\frac{
\braket{\hat \Pi_z^2}{\varepsilon}
-
\braket{\hat \Pi_z}{\varepsilon}^2
}{
\left(
\frac{\partial}{\partial\phi}
\braket{\hat \Pi_z}{\varepsilon}
\right)^2
}
\notag\\
&=
\Delta^2_{\Pi_z,0}
+
\frac{
(1-2\varepsilon)^{-2N_S}
-
1
}{
\left(
\frac{\partial}{\partial\phi}
\braket{\hat \Pi_z}_{0}
\right)^2
}.
\label{eq:direct_parity_noise}
\end{align}
Thus detection noise multiplies the parity signal by an exponentially small contrast factor, which appears as an additional contribution to the estimator variance.

For the GHZ probe state $\ket{\psi_{\rm GHZ}}\propto\ket{\uparrow}^{\otimes N_S}+\ket{\downarrow}^{\otimes N_S}$, the parity signal oscillates at frequency $N_S$, so the noiseless slope satisfies
\begin{align}
\left(
\frac{\partial}{\partial\phi}
\braket{\hat \Pi_z}{0}
\right)^2
\le
N_S^2.
\end{align}
For small $\varepsilon$, the excess factor scales as
\begin{align}
    (1-2\varepsilon)^{-2N_S}-1=\mathcal O(\varepsilon N_S).
\end{align}
Therefore, preserving the Heisenberg-limited scaling $\Delta\phi^2\sim 1/N_S^2$ requires $\varepsilon=\mathcal O(1/N_S)$. Direct parity readout of large GHZ states thus demands increasingly high single-ion detection fidelity.

\subsection{QLS parity readout via logic detection}
\label{app:QLS_parity}

We now contrast this with parity readout through collective QLS. In the limit $\theta\to\pi$, Eq.\eqref{eq:parity_measurement} shows that the spectroscopy parity is encoded in the collective logic signal $\braket{{\hat L}_x}$. Detection noise therefore acts on the logic ensemble rather than directly on all spectroscopy ions. After rotating the logic $x$ basis to the detection basis, the channel acts on the relevant collective observables as
\begin{align}
\Lambda({\hat L}_x)
&=
(1-2\varepsilon){\hat L}_x,
\notag\\
\Lambda({\hat L}_x^2)
&=
(1-2\varepsilon)^2{\hat L}_x^2
+
\frac{N_L}{4}
\left[
1-(1-2\varepsilon)^2
\right]\mathds{1}.
\end{align}
Using $
\braket{{\hat L}_x}_{0}
=
\frac{N_L}{2}
(-1)^{N_S}
\braket{\hat \Pi_z}_{0}$,
the corresponding estimator variance becomes
\begin{align}
\Delta^2_{{\hat L}_x,\varepsilon}
=
\Delta^2_{\hat \Pi_z,0}
+
\frac{1}{N_L}
\frac{
(1-2\varepsilon)^{-2}
-
1
}{
\left(
\frac{\partial}{\partial\phi}
\braket{\hat \Pi_z}_{0}
\right)^2
}.
\label{eq:qls_parity_noise}
\end{align}

Equation\eqref{eq:qls_parity_noise} shows the key advantage of QLS-based parity detection. The detection-noise penalty no longer grows exponentially with the number of spectroscopy ions, because the fragile many-body parity is first mapped onto a collective logic observable. As a result, maintaining Heisenberg-limited sensitivity does not require the detection error probability $\varepsilon$ to decrease with $N_S$. Instead, for fixed $\varepsilon$, the residual detection-noise contribution can be suppressed by increasing the number of logic ions, $N_L$.

\subsection{Mode-frequency fluctuations}
\label{app:mode_detuning}

We now analyze the effect of a finite detuning $\delta$ between the spin-dependent-displacement drives and the motional mode. The QLS unitary becomes

\begin{widetext}

\begin{align}
\hat U
&=
e^{-i t_L \left(\delta \hat a^{\dagger}\hat a-\frac{g_L}{\sqrt N}{\hat L}_z(\hat a+\hat a^{\dagger})\right)}
e^{-i t_S \left(\delta \hat a^{\dagger}\hat a+i\frac{g_S}{\sqrt N}S_{\downarrow}(\hat a-\hat a^{\dagger})\right)}
e^{-i t_L \left(\delta \hat a^{\dagger}\hat a+\frac{g_L}{\sqrt N}\hat L_z(\hat a+\hat a^{\dagger})\right)}
\label{eq:detuned_unitary}
\end{align}

Using the convention
$
D(\hat\alpha)=\exp\!\left(\hat{\alpha} \hat a^\dagger-\hat{\alpha}^{\dagger}\hat a\right),
$
the unitary can be rewritten into 
\begin{align}
\hat U
&=
e^{-i\Theta_L\hat L_z^2}
e^{-i\Theta\hat L_z\hat S_\downarrow}
e^{-i\Theta_S\hat S_\downarrow^2}
D\!\left(\chi_L\hat L_z+\chi_S\hat S_\downarrow\right)
e^{-i\delta(2t_L+t_S)\hat a^\dagger\hat a},
\end{align}
where $\Theta_L
=
-\frac{2g_L^2}{N\delta^2}
\left[
\delta t_L-\sin(\delta t_L)
-\left(1-\cos(\delta t_L)\right)
\sin\!\left(\delta(t_L+t_S)\right)
\right], 
\Theta
=
-\frac{8g_Lg_S}{N\delta^2}
\sin\!\left(\frac{\delta t_L}{2}\right)
\sin\!\left(\frac{\delta t_S}{2}\right)
\cos\!\left(\frac{\delta(t_L+t_S)}{2}\right),\\
\Theta_S
=
-\frac{g_S^2}{N\delta^2}
\left[
\delta t_S-\sin(\delta t_S)
\right]$,
and
$
\chi_L
=
\frac{g_L}{\sqrt{N}\,\delta}
\left(1-e^{i\delta t_L}\right)
\left(e^{-i\delta(t_L+t_S)}-1\right),
\chi_S= 
-i\,\frac{g_S}{\sqrt{N}\,\delta}
\left(1-e^{i\delta t_S}\right)
e^{-i\delta t_L}$.
\end{widetext}
The term proportional to $\hat S_{\downarrow}^2$ acts only on the spectroscopy ensemble and commutes with logic observables, while the final free motional evolution does not affect measurements on the logic ions and therefore both can be neglected when calculating the logic ion expectation values. The residual displacement, however, leaves logic and motion imperfectly disentangled. For logic-only observables, the relevant contribution is the logic-dependent part $\chi_L {\hat L}_z$; the spectroscopy-dependent displacement cancels from the logic-spin expectation values.

For an initial thermal state of the motional mode,
\begin{align}
\hat \rho_{\rm th}
=
\sum_{n=0}^{\infty}
\frac{\bar n^n}{(\bar n+1)^{n+1}}
\ket{n}\bra{n},
\end{align}
with mean occupation $\bar n$, the logic-spin moments become
\onecolumngrid
\begin{align}
\braket{{\hat L}_x}
&=
\frac{N_L}{2}
\cos^{N_L-1}(\Theta_L)
e^{-|\chi_L|^2/2(2\bar n+1)}
\sum_{n_\downarrow}
p_{n_\downarrow}
\cos(\Theta n_\downarrow),
\label{eq:Lx_detuned}
\end{align}
and
\begin{align}
\braket{{\hat L}_x^2}
&=
\frac{N_L(N_L+1)}{8}
+
\frac{N_L(N_L-1)}{8}
\cos^{N_L-2}(2\Theta_L)
e^{-2|\chi_L|^2(2\bar n+1)}
\sum_{n_\downarrow}
p_{n_\downarrow}
\cos(2\Theta n_\downarrow).
\label{eq:Lx2_detuned}
\end{align}
\twocolumngrid

The exponential factors describe the loss of contrast due to residual logic-motion entanglement, while the factors involving $\Theta_L$ arise from the detuning-induced one-axis twisting of the logic ensemble.

We now assume that $\delta$ fluctuates between experimental repetitions according to a Gaussian distribution with zero mean and variance $\sigma_\delta^2$. Since the instantaneous detuning is not known before each run, the pulse durations are chosen according to the resonant condition, and the detuning enters perturbatively. We expand Eqs.~\eqref{eq:Lx_detuned} and \eqref{eq:Lx2_detuned} to second order in $\delta$, replace $\delta^2$ by $\sigma_\delta^2$, and evaluate the estimator variance using the disorder-averaged moments. For simplicity we set $t_L=t_S/2$.

In the small-interaction regime, with pulse times chosen such that the resonant interaction parameter is $\pi/N_S$, the result at $\phi=\pi/2$ is
\onecolumngrid
\begin{align}
\Delta^2_{\phi}=\frac{4}{\pi^{2}N_L}
+
\frac{1}{N_S}
\left[
1+
\frac{9(1+2\bar n)}{4g_S^{2}}\,\sigma^{2}_{\delta}
+\mathcal{O}\!\left(\sigma^{4}_{\delta}\right)
\right]
+\mathcal{O}\!\left(N_S^{-2}\right)
\label{eq:detuning_small_transfer}
\end{align}
\twocolumngrid
Thus SQL scaling is retained if $N_L=\mathcal O(N_S)$ and the detuning-induced correction remains perturbative. In particular, it is sufficient that $\sigma_\delta^2(2\bar n+1)/(g^2_S)$ does not grow with system size and remains small.

In the large-interaction regime, with pulse times chosen such that the resonant interaction parameter is $\pi$, one obtains for a spectroscopy coherent spin state
\onecolumngrid
\begin{align}
\Delta_{\phi}^2=
\frac{\cos^{2-2N_S}(\phi)-\cos^2(\phi)}
{N_S^2\sin^2(\phi)}
+
\frac{
\pi^2N
\left[
32g_L(N_L-1)\pi
+
81g_S(1+2\bar n)
\right]
\cos^{2-2N_S}(\phi)
}{
144g_S^3N_LN_S^2\sin^2(\phi)
}\sigma_{\delta}^2
+
\mathcal{O}(\sigma_{\delta}^4)\label{eq:detuning_large_transfer_css}
\end{align}

\twocolumngrid
The detuning correction diverges as $\phi\rightarrow 0$, so the protocol must be operated at a finite working point. The most favorable scaling is obtained for $\phi\sim 1/\sqrt{N_S}$. In this regime, and for a fixed number of logic ions, $N_L=\mathcal{O}(1)$, one has
$
\cos^{2-2N_S}(\phi)=\mathcal{O}(1),
\qquad
\sin^2(\phi)=\mathcal{O}(1/N_S).
$
Consequently, the ideal contribution in Eq.~\eqref{eq:detuning_large_transfer_css} retains SQL scaling,
$
\frac{\cos^{2-2N_S}(\phi)-\cos^2(\phi)}
{N_S^2\sin^2(\phi)}
=
\mathcal{O}(1/N_S).
$
Moreover, since $N=N_S+N_L=\mathcal{O}(N_S)$ for fixed $N_L$, the leading detuning correction scales as
$
\frac{\sigma_\delta^2}{g_S^2}
\left[
1+2\bar n
+
\frac{g_L}{g_S}(N_L-1)
\right],
$
up to numerical factors of order unity. In particular, for a single logic ion, $N_L=1$, the contribution proportional to $g_L$ vanishes, and the correction reduces to
$
\frac{\sigma_\delta^2}{g_S^2}(1+2\bar n).
$
Therefore, preserving SQL scaling requires
$
\frac{\sigma_\delta^2}{g_S^2}
\left[
1+2\bar n
+
\frac{g_L}{g_S}(N_L-1)
\right]
=
\mathcal{O}(1/N_S).
$

For comparison, consider a GHZ spectroscopy state in the parity-readout regime.
The ideal estimator variance scales as $1/N_S^2$. In the presence of detuning fluctuations, the sensitivity is best away from the zeros of $\sin(N_S\phi)$, the detuning correction is minimized when $\sin^2(N_S\phi)$ is maximal. This occurs, for example, at $\phi=\pi/(2N_S)$, where $\sin^2(N_S\phi)=1$. At this operating point the ideal contribution scales as
\onecolumngrid
\begin{align}
\Delta_{\phi}^2=
\frac{1}{N_S^{2}}
+
\frac{\pi^{2}(N_L+N_S)\left[32g_L(N_L-1)\pi+81g_S(1+2\bar n )\right]}
{144\,N_L\,N_S^{2}\, g_S^{3}}\sigma_{\delta}^{2}+\left(
\frac{\pi^4 N_S\left(1+2N_L+N_S\right)}
{12g_L^2g_S^2N_L}+\mathcal{O}(N_S^0)\right)\sigma_{\delta}^{4}
+\mathcal{O}\!\left(\sigma_{\delta}^{6}\right)
\label{eq:detuning_large_transfer_ghz}
\end{align}
\twocolumngrid
Heisenberg scaling requires each detuning-induced correction in Eq.~\eqref{eq:detuning_large_transfer_ghz} to scale at most as $N_S^{-2}$. The coefficient of $\sigma_\delta^2$ scales as $N_S^{-1}$, whereas the coefficient of $\sigma_\delta^4$ scales as $N_S^2$. The quadratic term therefore requires $\sigma_\delta^2=\mathcal{O}(N_S^{-1})$, while the quartic term imposes the stronger condition $\sigma_\delta^2=\mathcal{O}(N_S^{-2})$. For $N_L\sim N_S$, the coefficient of $\sigma_\delta^2$ again scales as $N_S^{-1}$, since both $N_L+N_S$ and the term proportional to $g_L(N_L-1)$ are $\mathcal{O}(N_S)$, while the denominator is $\mathcal{O}(N_S^3)$. In this regime, however, the coefficient of $\sigma_\delta^4$ scales as $N_S$, so the quartic contribution requires $\sigma_\delta^2=\mathcal{O}(N_S^{-3/2})$. Thus, through quartic order, Heisenberg scaling is preserved provided that $\sigma_\delta^2=\mathcal{O}(N_S^{-2})$ for $N_L=\mathcal{O}(1)$ and $\sigma_\delta^2=\mathcal{O}(N_S^{-3/2})$ for $N_L\sim N_S$.

\section{Beyond the Lamb--Dicke approximation}
\label{app:LambDickeCorrections}

Throughout the main text we work in the Lamb--Dicke approximation. Here we estimate the leading corrections by retaining the first phonon-number-dependent contribution to the sideband coupling. For the logic and spectroscopy spin-dependent displacements we replace Eqs.\eqref{eq:H_L} and \eqref{eq:H_S} by
\begin{align}
\hat H_S
&=\frac{g_S}{\sqrt N}
\hat S_\downarrow i
\left[
\hat a -\hat a ^\dagger
-
\zeta\frac{\eta^2}{2N}
\left(
\hat a{\hat a}^\dagger {\hat a}-{\hat a}^\dagger {\hat a}\hat a^\dagger
\right)
\right],\\
\hat H_L
&=
\frac{g_L}{\sqrt N}
\hat L_z
\left[
\hat a+\hat a^\dagger
-
\frac{\eta^2}{2N}
\left(
\hat a{\hat a}^\dagger {\hat a}+{\hat a}^\dagger {\hat a}\hat a^\dagger
\right)
\right]
\label{eq:LD_corrected_hamiltonians}
\end{align}
Here $\eta$ is the single-ion Lamb--Dicke parameter of the logic ions and $\zeta=\eta_S^2/\eta_L^2$ accounts for a possible imbalance between spectroscopy and logic Lamb--Dicke parameters.

The full QLS sequence is
\begin{align}
e^{-i\theta \hat {\mathcal{H}}}
=
e^{i\hat H_Lt_L}
e^{-i\hat H_St_S}
e^{-i\hat H_Lt_L}.
\end{align}
Expanding the effective generator to order $\eta^2$ gives
\onecolumngrid
\begin{align}
\hat {\mathcal{H}}
=&
\hat S_\downarrow
\left[
\left(
\frac{1}{2\gamma_L}
-
\frac{1+3\zeta}{4}
\gamma_L
\frac{\eta^2}{N}
\hat L_z^2
\right)
i(\hat a-\hat a^\dagger)
-
\frac{\zeta\eta^2}{4N\gamma_L}
i
\left(
\hat a{\hat a}^\dagger {\hat a}-{\hat a}^\dagger {\hat a}\hat a^\dagger
\right)
\right]
\notag\\
&
+
\hat L_z \hat S_\downarrow
\left[
1
-
\frac{\zeta+1}{2N}
\eta^2
(2{\hat a}^\dagger {\hat a}+1)
+
\frac{\zeta-1}{4N}
\eta^2
(\hat a^2+\hat a^{\dagger 2})
-
\frac{1}{2}
\left(
\zeta+\frac{1}{3}
\right)
\gamma_L^2
\frac{\eta^2}{N}
\hat L_z^2
\right]
+
\mathcal O(\eta^4),
\label{eq:O_beyond_LD}
\end{align}
\twocolumngrid
where $\gamma_L
=
\frac{g_Lt_L}{\sqrt N}, \gamma_S
=
\frac{g_St_S}{\sqrt N}, \theta
=
2\gamma_L\gamma_S$.

In the following we focus on the experimentally relevant case $\eta_S=\eta_L$, so that $\zeta=1$. In this case the pair-creation and pair-annihilation terms proportional to $\hat a^2+\hat a^{\dagger 2}$ vanish. The remaining spectroscopy-only displacement terms do not affect logic observables at order $\eta^2$. After separating the logic-dependent part of the evolution, the QLS sequence can be written as
\onecolumngrid
\begin{align}
\hat U
=
D
\left[
-\gamma_S \hat S_\downarrow
+
\eta^2\theta \hat L_z\hat S_\downarrow
\left(
\gamma_L\hat L_z
-
i\gamma_S \hat S_\downarrow
\right)
\right]
\exp
\left[
-i\theta \hat L_z\hat S_\downarrow
-
i\eta^2\theta \hat L_z\hat S_\downarrow
\left(
\frac{1}{3}\gamma_S^2\hat S_\downarrow^2
-
\frac{2}{3}\gamma_L^2\hat L_z^2
-
2{\hat a}^\dagger {\hat a}
-
1
\right)
\right]
+
\mathcal O(\eta^4).
\label{eq:LD_effective_unitary}
\end{align}
\twocolumngrid
The displacement in Eq.\eqref{eq:LD_effective_unitary} does not contribute to logic-spin moments at order $\eta^2$: its leading term is independent of the logic state, while the logic-dependent correction is order $\eta^2$ and therefore affects contrast only at order $\eta^4$. Thus the leading effect on logic observables comes from the second exponential. It consists of a phonon-number-dependent correction to the interaction parameter and nonlinear spin terms proportional to $\hat L_z^3 \hat S_\downarrow$ and $\hat L_z\hat S_\downarrow^3$.

Using Eq.~\eqref{eq:LD_effective_unitary}, and assuming a thermal motional state with mean occupation $\bar n$, one obtains the following corrections to the estimator variance in the regimes discussed in the main text. In the small-interaction regime, $\theta=\pi/N_S$, evaluated at $\phi=\pi/2$,
\onecolumngrid

\begin{align}
\Delta^2_{\phi}
=&
\frac{4}{N_L\pi^2}
+
\frac{1}{N_S}
+
\frac{\eta^2}{N_L(N_L+N_S)\pi^2}
\Bigg[
\frac{\pi}{N_S}
\left(
\frac{g_S}{g_L}
\left[
4N_S(N_S+1)
-
\pi^2
-
\frac{8}{3}
-
\frac{\pi^4}{8}
\right]
+
\frac{g_L}{g_S}
\frac{3N_L-2}{3}
\right)
+
8(1+2\bar n)
\Bigg]
\notag \\
&+
\mathcal O(\eta^4)
+
\mathcal O\left(\frac{1}{N_S^2}\right), 
\label{eq:LD_small_transfer}
\end{align}
where we have also used $t_S=2t_L$. 

\twocolumngrid
This expression preserves SQL scaling when $N_L=\mathcal O(N_S)$, provided that the single-ion Lamb–Dicke parameter remains constant as the number of ions in the crystal increases. Although the thermal contribution does not alter the asymptotic SQL scaling as long as $\bar n$ grows no faster than linearly with system size, it can nevertheless produce a substantial finite-size correction. In particular, when $\eta^2\bar n\gtrsim1$, thermal effects are no longer perturbatively small and can significantly degrade performance, especially for small $N_L$ and $N_S$. In this regime, the magnitude of the correction should therefore be calculated explicitly for the relevant experimental parameters rather than inferred solely from its asymptotic scaling.

In the large-interaction regime the order-$\eta^2$ correction vanishes exactly at $\theta=\pi$. To characterize deviations from this point, we set $\theta=\pi-\tilde\theta$ and expand for small $\phi$ and small $\tilde\theta$. For a spectroscopy coherent spin state,
\onecolumngrid

\begin{align}
\Delta^2_{\phi}
=&
\frac{1}{N_S}
+
\frac{\eta^2}{N_L+N_S}
\frac{\pi\phi\tilde\theta}{24N_LN_S}
\left[
\pi
\left(
8\frac{g_S}{g_L}
+
\frac{g_L}{g_S}
\frac{9N_L^2-9N_L+2}{4}
\right)
+
6(3N_L-1)(1+2\bar n)
\right]\notag \\
& +
\mathcal O(\eta^4)
+
\mathcal O(\phi^2)
+
\mathcal O(\tilde\theta^2).
\label{eq:LD_large_transfer_css}
\end{align}
\twocolumngrid
Away from this point the correction remains perturbative provided the product $\eta^2\phi\tilde\theta$ is sufficiently small. In particular SQL scaling is maintained if $
\eta^2 \phi \tilde\theta
\left(
1+2\bar n
\right)
\lesssim
N_S
$, given that $N_L=\mathcal{O}(1).$

Finally, for a GHZ spectroscopy state in the parity-readout regime one finds
\onecolumngrid

\begin{align}
\Delta^2_{\phi}
=&
\frac{1}{N_S^2}
+
\frac{\eta^2\tilde\theta}{N_S+N_L}
\frac{\pi(1+N_L)(1+N_S)}{24N_LN_S}
\frac{1}{\sin^2(N_S\phi)}
\left[
\pi\frac{g_S}{g_L}
\left(
N_S^2
+
5N_S
-
2
\right)
+
\frac{\pi}{4}\frac{g_L}{g_S}
(3N_L-2)
+
12\bar n
+
6
\right]\notag \\
&
+
\mathcal O(\eta^4)
+
\mathcal O(\tilde\theta^2).
\label{eq:LD_large_transfer_ghz}
\end{align}

\twocolumngrid

For $N_L=\mathcal{O}(1)$ the requirement to maintain Heisenberg scaling is $
\eta^2\tilde\theta=\mathcal{O}(1/N_S^3)$, which means that implies that under this condition the requirement on thermal bosons is only $\bar{n}\lesssim N_S^2$. 

\section{Role of the initial logic--motion entangling step}
\label{app:initial_logic_motion}

In this appendix we show why the initial logic--motion entangling step in the QLS sequence is essential. We compare the protocol used in the main text to a simplified variant in which the spectroscopy-dependent displacement is followed directly by a logic-dependent displacement, without the initial logic--motion displacement and its time reversal. The corresponding unitary is
\begin{align}
e^{-it_LH_L}
e^{-it_SH_S}
e^{i\theta L_zS_\downarrow/2}
D
\left(
-i\gamma_LL_z-\gamma_SS_\downarrow
\right),
\label{eq:simple_protocol_unitary}
\end{align}
where $\gamma_L
=
\frac{g_Lt_L}{\sqrt N}, 
\gamma_S
=
\frac{g_St_S}{\sqrt N}, 
\theta
=
2\gamma_L\gamma_S$.
Thus the simplified sequence produces a spectroscopy-dependent rotation of the logic ions, but it also leaves behind a logic-dependent displacement of the motional mode. This residual displacement entangles the logic ions with the motion and reduces the logic measurement contrast.

For an initial thermal motional state with mean occupation $\bar n$, the relevant overlap of the displaced motional states is $
\left[\rho_{\rm th}
D
\left(
-i\gamma_L(m-m')
\right)\right]
=
\exp
\left[
-\frac{1}{2}
\gamma_L^2(m-m')^2
(2\bar n+1)
\right]$.
Since $L_x$ connects logic Dicke states with $m-m'=\pm1$, this gives
\begin{align}
\braket{L_x}
=
\frac{N_L}{2}
e^{-\gamma_L^2(2\bar n+1)/2}
\sum{n_\downarrow=0}^{N_S}
p_{n_\downarrow}
\cos(\theta n_\downarrow).
\label{eq:Lx_simple_protocol}
\end{align}
Similarly, because $L_x^2$ contains coherences with $m-m'=\pm2$, one obtains
\onecolumngrid
\begin{align}
\braket{L_x^2}
=
\frac{N_L(N_L+1)}{8}
+
\frac{N_L(N_L-1)}{8}
e^{-2\gamma_L^2(2\bar n+1)}
\sum_{n_\downarrow=0}^{N_S}
p_{n_\downarrow}
\cos(2\theta n_\downarrow).
\label{eq:Lx2_simple_protocol}
\end{align}
\twocolumngrid
Compared with Eqs.\eqref{eq:Jx_general} and \eqref{eq:Jx2_general}, the simplified protocol therefore has the same dependence on the spectroscopy population distribution, but the logic signal and its second moment are suppressed by temperature-dependent contrast factors. These factors arise from uncompensated logic--motion entanglement.

Using Eqs.\eqref{eq:Lx_simple_protocol} and \eqref{eq:Lx2_simple_protocol}, the estimator variance in the small-interaction regime, evaluated at $\phi=\pi/2$ and $\theta=\pi/N_S$, becomes
\onecolumngrid
\begin{align}
\Delta^2_{\phi}
=
&\frac{4}{N_L\pi^2}
\left[
\cosh A
+
N_L\sinh A
\right]
+
\frac{1}{N_S}
\left[
\cosh A
+
\frac{1}{N_L}
\sinh A
\right]
+
\mathcal O
\left(
\frac{1}{N_S^2}
\right), 
\label{eq:simple_protocol_small_transfer}
\end{align}
\twocolumngrid
where $A=(1+2\bar n)\gamma_L^2$
Thus, even in the small-interaction regime, the residual logic--motion entanglement adds a contrast penalty that grows with both $\gamma_L^2$ and the thermal occupation.

In the large-interaction regime, taking $\theta\rightarrow\pi$, one finds
\onecolumngrid
\begin{align}
\Delta_\phi^2
=
\frac{
-N_L\cot^2(\phi)
+
\cos^{-2N_S}(\phi)
\cot^2(\phi)
\left[
N_L\cosh A
+
\sinh A
\right]
}{
N_LN_S^2
}.
\label{eq:simple_protocol_large_transfer}
\end{align}
\twocolumngrid
This expression reduces to the ideal result only when $A\rightarrow0$. For any finite $\gamma_L^2(2\bar n+1)$, the residual spin--motion entanglement degrades the signal-to-noise ratio and makes the protocol explicitly sensitive to the motional temperature.

These results show that the initial logic--motion entangling step is not merely a technical complication. It is the operation that allows the final inverse displacement to cancel residual logic--motion entanglement, thereby removing the temperature-dependent contrast factors in Eqs.~\eqref{eq:Lx_simple_protocol} and \eqref{eq:Lx2_simple_protocol}. The full time-reversal sequence analyzed in the main text is therefore essential for maintaining thermal robustness and generally yields higher measurement contrast, and hence improved sensitivity, than a sequence that omits the initial logic--motion entangling step.

\bibliography{library}

\end{document}